\documentclass[submission]{SciPost}
\pdfoutput=1

\usepackage{float}
\restylefloat{table}
\usepackage{microtype}
\usepackage{xcolor}
\usepackage{bbm}
\usepackage[utf8]{inputenc}
\usepackage{lmodern}
\usepackage{amsmath,amsfonts,amssymb,mathtools,bbold,mathrsfs}
\usepackage{babel}
\usepackage[sort&compress,numbers]{natbib}
\usepackage{tocloft}

\definecolor{yellow1}{HTML}{ffffcc}
\definecolor{myviolet}{HTML}{007B8B}
\definecolor{mylavender}{HTML}{DEFEFF}
\definecolor{soft}{HTML}{fff6ed}
\definecolor{sky}{HTML}{CEFFFF}
\definecolor{indigo}{HTML}{000066}
\definecolor{myblue}{HTML}{056FB1}
\definecolor{maroon}{HTML}{db0000}
\definecolor{forest}{HTML}{06961c}
\definecolor{lime}{HTML}{fcffad}
\definecolor{mygreen}{HTML}{024f16}
\definecolor{newgreen}{HTML}{cf5600}
\usepackage{hyperref}
\usepackage{prettyref}
\usepackage{soul}
\newcommand{\pref}{\prettyref}

\newrefformat{eq}{equation {(\ref{#1})}}
\newrefformat{app}{appendix {\ref{#1}}}
\newrefformat{sec}{section {\ref{#1}}}
\renewcommand{\d}{\mathrm{d}}
\newcommand{\ads}{ $AdS_5 \times S^5$ }

\hypersetup{
	colorlinks=true,
    urlcolor=blue,
	linkcolor=blue,
	filecolor=cyan,      
	citecolor=red
}

\begin{document}
\begin{samepage}
		\begin{flushleft}{ \huge \textbf{On-Shell Action for Type IIB Supergravity and Superstrings on \ads}}\end{flushleft}
		\vspace{20pt}
		{\color{myviolet}\hrule height 1mm}
		
		\vspace*{10pt}
		\begin{flushleft}
		\large	\textbf{Subhroneel Chakrabarti}$\,{}^a$, \textbf{Divyanshu Gupta}$\,{}^b$, \textbf{Arkajyoti Manna}$\,{}^c$
		\end{flushleft}

		\begin{flushleft}
			\emph{\large ${}^a$ FZU - Institute of Physics of the Czech Academy of Sciences \& CEICO, Na Slovance 2, 182 21 Prague 8, Czech Republic. \\ \vspace*{0.1cm}
            ${}^b$ Institute of Physics, University of Amsterdam, Science Park 904, 1098XH Amsterdam, The Netherlands. \\ \vspace*{0.1cm}
            ${}^c$ Center for High Energy Physics, Indian Institute of Science, C.V. Raman Avenue, Bangalore 560012, India.
			\\ \vspace{3mm}
			\href{mailto:subhroneelc@fzu.cz}{subhroneelc@fzu.cz}, \href{mailto:divyanshu.gupta@student.uva.nl}{divyanshu.gupta@student.uva.nl},  \href{mailto:arkajyotim@iisc.ac.in}{arkajyotim@iisc.ac.in}}\\
		\end{flushleft}
		

		\section*{Abstract}
		{\bf
			AdS/CFT predicts that the value of the on-shell action for type IIB Supergravity (SUGRA) on \ads background must be a non-zero number completely determined from the boundary theory. We examine this statement within Sen's formalism for type IIB SUGRA and find that consistency with AdS/CFT requires us to add a specific boundary term to the action. We contrast our resolution with two other resolutions recently proposed in the literature in the context of different approaches to type IIB SUGRA. We explain how our resolution presents a strong benchmark for the possible boundary term of the complete spacetime action for type IIB superstring and how it may possibly lead to a piece of evidence for the strongest form of AdS/CFT conjecture in \ads. We also comment on the fate of the on-shell action for general self-dual $p$-form fields in Sen's formalism in any curved backgrounds.
		}
		
	\end{samepage}

\newpage
	\vspace{10pt}
	\noindent\rule{\textwidth}{1pt}
	\pagecolor{white}
	\tableofcontents\thispagestyle{fancy}
	\noindent\rule{\textwidth}{1pt}
	\vspace{10pt}

\section{Introduction} \label{sec:intro}

The type IIB supergravity in $10$-dimensions has famously three known maximally super-symmetric vacua - two of which ($10$d Flat spacetime and PP-wave background) can be interpreted as a limit of the third, the celebrated $AdS_5 \times S^5$ \citep{Maldacena:1997re, Berenstein:2002jq}. The construction of a Lorentz covariant action for type IIB SUGRA has been historically a challenge due to the presence of a self-dual RR 5-form flux. On the other hand, the equations of motion of the type IIB SUGRA on any background are unambiguously known. Ergo, on \ads these EOM can be dimensionally reduced to effective $5$d equations (along the $AdS_5$), which can be interpreted as Euler-Lagrange equations of a $5$d effective Lagrangian. This effective action obtained without invoking a parent action in $10$d is also consistent with AdS/CFT expectations. 

The field content (bosonic)\footnote{Throughout this paper, we focus only on bosonic parts of the field content. The fermionic parts are fixed completely from supersymmetry once the answers are known for the bosonic part.} of type IIB SUGRA is as follows \citep{,SCHWARZ1983301,HOWE1984181,SCHWARZ1983269}
\begin{itemize}
    \item The spacetime metric $g_{ab}$ ($a,b = 0,1, \cdots,9$).
    \item The dilaton $\phi$.
    \item The Kalb-Ramond 2-form $B^{(2)}$ whose field strength is denoted as $H^{(3)} = \d B^{(2)}$.
    \item The Ramond-Ramond fluxes (field strengths) - $F^{(1)}, F^{(3)}, F^{(5)}$. Among these the $5$-form is self-dual, i.e. $F^{(5)} = \star_g F^{(5)}$.
\end{itemize}

The solutions of these fields that give us \ads background are given as (throughout this paper, we use $\cong$ to mean ``equal on-shell'')
\begin{align} \label{eq:Ads5}
d s_{10}^{2}& = g_{ab} \; \d x^a \d x^b \nonumber \\
&\cong \rho^{2}\left(d s_{\mathrm{AdS}_{5}}^{2}+d \Omega_{5}^{2}\right) , \nonumber \\
F^{(5)} & \cong 4 \rho^{-1}\left(\epsilon_{5}+ \star_g  \epsilon_{5}\right) \nonumber\\
B^{(2)} \cong F^{(1)} \cong F^{(3)} \cong \partial_a \phi &\cong 0  \;.
\end{align}
Here, the line element of $AdS_5$ and the $S^5$ are denoted by $d s_{\mathrm{AdS}_{5}}^{2}$ and $d \Omega_{5}^{2}$ respectively, $\epsilon_{5}$ is the epsilon tensor in $AdS_5$, $\rho$ is the radius of $\text{AdS}_5$ space given by $\rho^{4} =4 \pi \alpha^{\prime 2} g_{s} N$, $\alpha'$ is the Regge slope parameter, $g_s$ is the string coupling,  and $N$ corresponds to the units of RR 5-form flux in the background geometry.

The 5d effective action evaluated in this background gives 
\begin{equation} \label{eq:5d_onshell}
S_{5} \cong \frac{8 \rho^{4}}{2 \kappa_{5}^{2}} \operatorname{vol}\left(\operatorname{AdS}_{5}\right) \;.
\end{equation}
Here $\kappa_5$ is Newton's constant in $5$d. Alternatively, in terms of the 10d gravitational constant, we have 
\begin{equation}
   S_{5} \cong \frac{4 \rho^8}{\kappa_{10}^2} \operatorname{vol}\left(\mathrm{AdS}_5\right) \;,\; \text{where }\, \frac{1}{2 \kappa_5^2}=\frac{\rho^4 \operatorname{vol}\left(S^5\right)}{2 \kappa_{10}^2} \; \;.
\end{equation}
Here we are deliberately using the same convention for the $5$d on-shell action as \citep{Kurlyand:2022vzv} for ease of comparison later. This is indeed consistent with AdS/CFT, which predicts the 5d on-shell action must be given by the conformal a-anomaly of the $\mathcal{N}=4,\, SU(N)$ Super-Yang-Mills theory living on the boundary $S^4$ \citep{Liu:1998bu, Henningson:1998gx, Gubser:1998vd, Blau:1999vz, Burgess:1999vb, Russo:2012ay}. 

The Ricci scalar for the \ads metric can be checked to be vanishing for the full 10d, i.e.,  $R \cong 0$ (the $AdS_5$ and $S^5$ spaces have exactly equal and opposite scalar curvature). This means that along with the Einstein-Hilbert term, if one writes the usual Maxwell-like term for the only non-zero field strength as the action for type IIB SUGRA, then it will trivially vanish on the solution.

But this vanishing does not make use of the details of the solution for the field strength. For any self-dual form, i.e. $F = \star_g F$, the Maxwell action $F \wedge \star_g F = F \wedge F $ identically vanishes irrespective of the exact form of the field strength! In fact, this is precisely the obstacle that prevented the community from writing down a manifestly Lorentz covariant action for type IIB SUGRA. The most commonly adopted workaround in the literature has been to instead work with a \emph{pseudoaction} that uses a Maxwell-like term and imposes the self-duality constraint \emph{by hand} on the equations of motion. While the pseudoaction is perfectly reasonable to work with in obtaining the EOM, it, by construction, is not the correct action for the type IIB SUGRA.

The vanishing of the on-shell 10d pseudoaction and the non-vanishing of the on-shell 5d effective action is presented as a puzzle in \citep{Kurlyand:2022vzv}. It was argued there that this puzzle requires resolution even within the framework of pseudoaction since the quantity in question is an on-shell one. 

We can extend the puzzle as follows - any claim for the correct action for type IIB SUGRA must give a non-zero on-shell value on \ads, which is consistent with the AdS/CFT predictions once the sphere directions are integrated out. In fact, assuming the strongest form of the AdS/CFT conjecture, we can, in fact, claim that the spacetime action for full type IIB strings (i.e. type IIB string field theory action \citep{Sen:2015uaa}) must on-shell give a non-zero value consistent with AdS/CFT prediction. 

In this paper, we investigate and resolve the aforementioned puzzle by working with the action for type IIB SUGRA given by Sen in \citep{Sen:2015nph}. Our resolution for this puzzle echoes the core principle of the proposal in \citep{Kurlyand:2022vzv}, i.e. the type IIB SUGRA action must be supplemented with a suitable boundary term, which will not affect the equations of motion but give a non-vanishing value to the on-shell $10$d action. In fact, while our proposed boundary term has a different structure off-shell, on-shell, it matches precisely with the corresponding answer in \citep{Kurlyand:2022vzv}.

Another recent paper in \citep{Mkrtchyan:2022xrm} has given a new action for type IIB SUGRA (in fact, the proposed action describes both type IIA and type IIB on the same footing). It was briefly discussed there how the proposed action for type II SUGRA already contains a term that on the \ads background for type IIB gives the expected answer. Given the result of the papers \citep{Kurlyand:2022vzv, Mkrtchyan:2022xrm}, one may wonder why anyone should try to resolve the puzzle again using a different action. We give three following reasons. 

\begin{itemize}
    \item Sen's action is naturally related to the spacetime action for superstrings, viz. type IIB SFT action  \citep{Sen:2015uaa}. Therefore we will see the resolution of this puzzle in Sen's formalism provides a benchmark for the structure of boundary terms for full string theory. In fact, in light of the recent result presented in \citep{Erler:2022agw}, any insight into structures of possible boundary terms in string field theory at this point is a considerable step forward. In contrast, it is not clear how the action proposed in \citep{Mkrtchyan:2022xrm} is embedded into the full string theory while \citep{Kurlyand:2022vzv} only works with the pseudoaction. 
   
    \item The resolution of the puzzle should not be dependent on a particular approach to describe type IIB SUGRA. While each approach has its relative advantages, all approaches must agree on every physical result. The on-shell action evaluated on the \ads background is one such physical result, and it needs to be solved within Sen's formalism independently.

     \item Sen's construction can be used to write down the action for any self-dual $2k+1$-form field strengths in $4k+2$ dimensions \citep{Sen:2019qit}. The result of this paper also leads to an understanding of possible boundary terms for other chiral $p$-form theories. Since Sen's construction prima facie is a little unusual, it is interesting in its own right to understand the structure of boundary terms within this formalism.  
\end{itemize}

The structure of this paper is as follows. In \pref{sec:review} we give a summary of Sen's formalism for any self-dual form field that will be needed to follow the main result of this paper. In \pref{sec:IIBSen} we give the action for type IIB SUGRA due to Sen and show that despite not having a Maxwell-like action, it still vanishes on-shell. We also show in this section that this is true for any chiral $p$-form described in this formalism. Therefore in \pref{sec:boundary}, we supplement Sen's action by a pure boundary term and show how it leads to a consistent result with AdS/CFT prediction. With the resolution at hand for type IIB SUGRA, we outline what can we learn about the resolution of the puzzle in full type IIB string theory in \pref{sec:SFT}. We conclude with some comments and a summary in \pref{sec:end}.

\section{A Quick Review of Sen's Formalism}
\label{sec:review}

In this section, we briefly review the string field theory inspired Lagrangian description for any self-dual $(2k+1)$-form field strength in $(4k+2)$ spacetime dimensions due to Sen \citep{Sen:2015nph,  Sen:2019qit}.  In this formalism, the self-duality condition holds off-shell, the action is polynomial, while preserving manifest Lorentz invariance at the cost of introducing a single additional unphysical field that completely decouples from the dynamics. Some recent works using this formalism in various dimensions at both classical and quantum level are \citep{Lambert:2019diy, Andriolo:2020ykk, Gustavsson:2020ugb, Chakrabarti:2020dhv, Andriolo:2021gen, Rist:2020uaa, Cremonini:2020skt, Chakrabarti:2022lnn, Andrianopoli:2022bzr}.

The action in this formulation contains a $2k$-form field $P$, a self-dual $(2k+1)$- form field strength $Q$ satisfying 
\begin{align}
\star Q=Q\,
\end{align}
 and a linear map $M(Q)$ that maps self-dual forms to anti-self-dual forms.  That is for all $\star Q= Q$, we have
\begin{align}
\star M(Q)=-M(Q)\,.
\end{align}
The Hodge star operation $\star$ is defined with respect to the flat metric and not with the actual background metric. The Hodge dual with respect to the dynamical metric will be denoted by $\star_g$.
See \citep{Sen:2015nph,Sen:2019qit,Andriolo:2020ykk} for more details on the explicit construction of the map $M(Q)$. We will not require its explicit form in this paper.

The action for the self-dual field takes the following form
\begin{align}
S=\frac{1}{2} \int d P \wedge \star \, dP - \int d P \wedge Q  + \int Q  \wedge M(Q) \label{eq:Sen'saction}\,.
\end{align}
  In the canonical formalism, it was shown in \citep{Sen:2019qit} that the Hamiltonian splits into a sum of a free Hamiltonian with only non-physical degrees of freedom and an interacting Hamiltonian containing only the physical degrees of freedom. The gravitational coupling to the dynamical metric enters \textit{only} through the map $M(Q)$, and the extra non-physical field does not couple even to gravity. The form field $P$ contributes solely to the non-physical field.

The invariance of the action in \pref{eq:Sen'saction} under diffeomorphisms is not manifest due to the non-standard coupling to the background metric. Nonetheless, the diffeomorphism symmetry of the action is indeed preserved, as shown in the original references \citep{Sen:2015nph,  Sen:2019qit}.

Due to the unusual self-duality condition on $Q$ and the presence of Hodge star in the kinetic term of $P$, the fields entering the action in \pref{eq:Sen'saction} are not standard differential forms on the background manifold. We refer to them as \textit{pseudoforms} following \citep{Andriolo:2020ykk}. Even though $Q$ and $M(Q)$ are individually not physical forms on the manifold, the following specific linear combination is a proper $(2k+1)$-form and satisfies the self-duality constraint w.r.t the background metric $g$
\begin{align}
Q-4M(Q)=\star _g(Q-4M(Q))\,.
\end{align}
This particular combination, on-shell, matches with the physical self-dual field strength obtained via a pseudoaction formalism. For our case, this implies the physical RR 5-form flux $F_{(5)}$ of type IIB SUGRA action in 10d will be given by the above combination \emph{on-shell}. \\

We conclude our discussion on Sen's action by emphasizing that this formalism does not need to evoke any notion of gauge potential anywhere. Rather everything (including possible interactions with other fields) is completely captured by the field-strength-like field. While this may seem unusual, as pointed out in the original paper \citep{Sen:2015nph}, this is a highly desirable feature from the string theory perspective, where it is known that the strings and the D-branes are only sensitive to the RR-fluxes and not the RR-potentials. We also like to reiterate that the action in \pref{eq:action} contains, along with the physical interacting degrees of freedom, a free, completely decoupled, unphysical extra field. For more details, we refer the reader to the original papers \citep{Sen:2015nph,Sen:2019qit} (also see \citep{Andriolo:2020ykk,Chakrabarti:2020dhv}).

\section{Sen's Action for Type IIB SUGRA} \label{sec:IIBSen}
For our purpose, it is enough to work with the bosonic part of the type IIB SUGRA action. Furthermore, since we are only interested in the on-shell value of the action for \ads background, it is sufficient to look into those terms of the action that do not trivially vanish once we go on-shell following \pref{eq:Ads5}. We will be following the normalization conventions of \citep{Sen:2019qit} along with the notation of \citep{Andriolo:2020ykk,Chakrabarti:2020dhv} for the gravitational coupling term. The relevant terms in the action are
\begin{equation} \label{eq:action}
S= S_{SD} + S_{EH} = \frac1{2} \int \d P \wedge \star \, \d P - \int \d P \wedge Q  + \int Q \wedge M(Q) + S_{EH} \;.
\end{equation}

 $S_{EH}$ is the Einstein-Hilbert action, which once again gives zero contribution on-shell since the Ricci scalar is zero for the \ads background.\\

This action gives the following equations of motion (we are not writing Einstein's field equations here)

\begin{equation} \label{eq:EOM}
\begin{aligned}
 \d \star \d P  - \d Q &=0 \\
\frac1{2} \left(\d P- \star \d P\right) + 2M(Q) &= 0 
\end{aligned}
\end{equation}

\subsection{On-Shell Action of Type IIB SUGRA In Sen's Formalism} \label{sec:On-shell}
We can make use of the equations of motion to obtain the on-shell value of the action given in \pref{eq:action}.

For example,
\begin{equation} \label{eq:2nd_term}
\begin{aligned}
- \int \d P \wedge Q  = -\frac1{2} \int (\d P- \star \d P) \wedge Q 
&\cong 2 \int  M(Q) \wedge Q = - 2 \int Q \wedge M(Q) \\
\implies -\int \d P \wedge Q + \int Q \wedge M(Q)  &\cong  -  \int Q \wedge M(Q)
\end{aligned}
\end{equation}

Similarly, we can do integration by parts of the kinetic term for the 4-form $P$ and write it as

\begin{equation} \label{eq:1st_term}
\begin{aligned}
\frac{1}{2} \int \d P \wedge \star \d P &= -\frac1{2} \int P \wedge \d \star \d P    
                                    \cong -\frac1{2} \int P \wedge  \d Q \\
 \implies  \frac{1}{2} \int \d P \wedge \star \d P &\cong \frac1{2}\int \d P \wedge Q \\
    \frac{1}{2} \int \d P \wedge \star \d P &\cong   \int Q \wedge M(Q)
\end{aligned}
\end{equation}

Plugging in the answer from \pref{eq:1st_term} and \pref{eq:2nd_term} into the action we get

\begin{equation} \label{eq:explicit_zero}
\begin{aligned}
     S_{SD} &= \frac{1}{2} \int \d P \wedge \star \d P -\int \d P \wedge Q + \int Q \wedge M(Q)\\ &\cong  \int Q \wedge M(Q) -  \int Q \wedge M(Q) \\ &\cong 0 \;.
\end{aligned}
\end{equation}
Therefore, we see that the part of Sen's action involving the self-dual field evaluated explicitly on-shell is identically zero.

It is curious that even though this action does not contain any Maxwell-like term, it still vanishes on-shell due to the fact that on-shell the contribution from the parts of the action depending on the field $Q$ exactly cancels the contribution from the part of the action independent of $Q$.

The rest of the terms in type IIB SUGRA all vanish on-shell since they are the same terms that appear in the pseudoaction. 

The result of \pref{eq:explicit_zero} actually holds for any self-dual form described by Sen's action in any appropriate dimensions. There is a shorter way of understanding why this is so, which we explain now.

\subsection{Why Sen's Action Vanishes On-shell?} \label{sec:why}
Consider the following (infinitesimal) transformation of the field variables $P$ and $Q$
\begin{equation}
\begin{aligned}
 P &\rightarrow(1+\epsilon)  P ,  \\
 Q &\rightarrow(1+\epsilon) Q ;  \quad \epsilon=\text { const. }, \; \epsilon <<1.
 \end{aligned}
\end{equation}
Under this transformation, the action transforms as
\begin{equation}
\begin{aligned} \label{eq:action_change}
    S[P+\epsilon P, Q+\epsilon Q] &= \frac1{2}(1+\epsilon) \int \d P \wedge \star \, \d P -(1+\epsilon) \int \d P \wedge Q + (1+\epsilon) \int Q \wedge M(Q) + O(\epsilon^2) \\
    &= (1 + \epsilon) S[P,Q] + O(\epsilon^2).
    \end{aligned}
\end{equation}
When evaluated on-shell, i.e. $P \cong P_0$ and $Q \cong Q_0$, the LHS of \pref{eq:action_change} gives back just $S[P_0,Q_0]$ due to the variational principle, and we have
\begin{equation} \label{eq:on-shell_zero}
    \begin{aligned}
       S[P_{0}, Q_{0}] &= (1+\epsilon) S[P_0, Q_0] \\
        \implies S[P_0,Q_0] &= 0 \;.
    \end{aligned}
\end{equation}
So, the on-shell action in Sen's formalism vanishes identically \textit{for any solution} of any self-dual $p$-form theory coupled to a non-dynamical curved background. For a dynamically curved background, Sen's action needs to be supplemented by the Einstein-Hilbert action and the usual Gibbons-Hawking-Yorke boundary term. These terms, while they did not contribute anything on-shell for type IIB on \ads may contribute to other self-dual theories coupled to dynamical gravity in other dimensions. 
\section{The Boundary Term} \label{sec:boundary}
It is now clear that to obtain a non-zero on-shell action we must supplement the action with a boundary term. Since on-shell, there is a clear link between Sen's action and the equations of motions obtained from the pseudoaction, it is tempting to think the proposal given in \citep{Kurlyand:2022vzv} will work here as well. However, the proposed term in \citep{Kurlyand:2022vzv} is a boundary term off-shell, if and only if we think of the RR 5-form flux variable as an exact form (i.e. a field strength of a gauge potential). As mentioned in \pref{sec:review} that this is not possible in Sen's action. So we need to look for a different resolution.

The boundary term we propose is
\begin{equation} \label{eq:boundary}
    S_b = \kappa \int \d \big[Q \, \wedge P \big] \;,
\end{equation}
where $\kappa$ is a constant which will be fixed by demanding consistency with AdS/CFT prediction.
Being a boundary term, clearly, this does not affect the equations of motion. To check how it indeed resolves the apparent mismatch, we need to find its on-shell value. It is a necessary evil of this formalism that any possible boundary term, even when evaluated on-shell, will contain a contribution from 3 types of terms. Firstly, terms which are purely built out of physical fields. Secondly, terms which are completely built out of the non-physical field. Finally, there can be terms which have a mixing of physical and the non-physical field. However, AdS/CFT correspondence is only cognizant of the physical field and nothing else. Therefore it stands to reason that only the completely physical part of the on-shell action in Sen's formalism should reproduce an answer consistent with holography.

Recall from \pref{sec:review}, that on-shell the self-dual RR 5-form flux is captured by the following specific linear combination
\begin{equation} \label{eq:physical}
    F^{(5)} \cong Q - 4 M(Q) \quad, \quad \d F^{(5)} \cong 0 \;.
\end{equation}

On-shell our boundary term, ergo, reduces to
\begin{equation}
    S_b \cong  \kappa \int P \wedge \d \star \d P + 2 \kappa \int Q \wedge M(Q) \;.
\end{equation}
Evidently, the first term is purely non-physical, and we can disregard it. Let us now focus on the second term. First of all, note that we can replace $Q$ by $F^{(5)}$ in the integrand. Next, we note that on the background of interest, the RR 5-form splits into a direct sum $F^{(5)} = F_{AdS} \oplus F_{S^5}$, with $F_{AdS} = \star_g F_{S^5}$. This allows us to write
\begin{equation}
     S_b \cong \kappa \int P \wedge \d \star \d P + 2\kappa \int F_{AdS} \wedge M(Q)_{S^5} +2 \kappa \int F_{S^5} \wedge M(Q)_{AdS} \;.
\end{equation}
Finally, we can use \pref{eq:physical} and the fact that $F_{AdS} = \star_g F_{S^5}$ to obtain

 \begin{equation}\label{eq:osb}
     S_b \cong \kappa \int P \wedge \d \star \d P + \frac{1}{2}\kappa \int F_{AdS} \wedge Q_{S^5} + \frac{1}{2}\kappa \int F_{S^5} \wedge Q_{AdS} +\frac1{2} \kappa \int F_{AdS} \wedge F_{S^5}  \;.
 \end{equation}

As discussed, the first term is purely built of the non-physical field, the second and third are of mixed type, and the final term is the purely physical part, which matches precisely with the proposed term in \citep{Kurlyand:2022vzv}. The constant $\kappa$ can now be fixed exactly along the line of \citep{Kurlyand:2022vzv} by comparing with \pref{eq:5d_onshell} (and using the solution for $F^{(5)}$ from \pref{eq:Ads5}). Also, note that we are using a normalization for the physical form field where a Maxwell-like term would be written as $\sim \frac{1}{5!} F \wedge \star_g F $. The value of the constant is
\begin{equation}\label{eq:kappa}
    \kappa =  \frac1{2 (5!)^2 \kappa_{10}^2} \;.
\end{equation}
Thus we see that the proposed boundary term makes the on-shell action non-zero, and it is indeed consistent with the AdS/CFT prediction. 

There are a few comments we want to make here. Firstly, on-shell, our proposal coincides precisely with the proposal of \citep{Kurlyand:2022vzv}. However, the off-shell realization is very different, which is expected since the pseudoaction, and Sen's action is inequivalent off-shell. Secondly, unlike \citep{Kurlyand:2022vzv}, our boundary term does not require a decomposition of the forms into the electric or the magnetic part, nor does it invoke the knowledge of the factorized nature of the background geometry prior to going on-shell. This feature of our proposed term is also shared by the formalism introduced in \citep{Mkrtchyan:2022xrm}. Third, in \citep{Kurlyand:2022vzv} it was shown that their boundary term follows from gauge invariance in the PST formulation of type IIB SUGRA \citep{Pasti:1996vs,DallAgata:1997gnw} and a similar observation was echoed in \citep{Mkrtchyan:2022xrm}. In contrast our approach required us to introduce this term by hand. This, in fact, is to be expected for Sen's formulation given its proximity to the complete spacetime action of type IIB string theory \citep{Sen:2015uaa, Sen:2015nph} (for a review see \citep{deLacroix:2017lif, Erler:2019loq, Erbin:2021smf}). In the next section, we will point out exactly what can be said about the on-shell spacetime action of the full type IIB string on \ads from our analysis. Finally, it is quite straightforward to see that our proposed boundary term will be equally valid for any other solution with the product structure of spacetime manifold, exactly like \citep{Kurlyand:2022vzv}. Once again, both proposals will agree on the value of the on-shell action.

\section{The on-shell Spacetime action of Type IIB String Theory on \ads} \label{sec:SFT}
Recently, it has been shown that the closed string field theory action must vanish on-shell up to possible boundary terms \citep{Erler:2022agw}. This is consistent with the known results for the effective action in spacetime obtained from worldsheet sigma models \citep{Fradkin:1984pq, Fradkin:1985ys, Callan:1986jb, Tseytlin:1988tv, Tseytlin:1988rr} How will the boundary terms in SFT look is an open question. On the other hand, there are known cases where on-shell actions in string theory are related to a non-zero quantity of physical interest, such as black hole entropy \citep{Susskind:1994sm, Chen:2021dsw}, D-brane tensions, and partition functions in dual matrix models \citep{Eberhardt:2021ynh, Mahajan:2021nsd}. These results strongly suggest that the closed SFT action for certain backgrounds needs to be supplemented with an appropriate boundary term. 

Given that the type IIB SUGRA action considered here was constructed with direct motivation from the closed type IIB SFT action (\citep{Sen:2015uaa}) and the \ads background is one for which the on-shell action must be non-zero (assuming the validity of AdS/CFT), the result obtained here provides a crucial benchmark for the possible boundary term that needs to be added to the type IIB SFT action in this case. 

 \ads is also expected to be an exact background for the complete type IIB superstring.  Therefore the only non-zero background fields for \ads solution for type IIB strings are also the metric and the RR $5$-form flux. Therefore, any string field on-shell must reduce to terms containing \textit{only} these two massless fields. Therefore, on-shell, even for the full string theory, the boundary term that resolves the puzzle is the same one we proposed. This also suggests that any possible boundary term in the full type IIB string field theory action must reduce to our proposal, up to possible field redefinitions, once it is evaluated on-shell. We can express this as (in the following $\mathbf{\Tilde{S}_{b}}$ is the boundary term for type IIB SFT action)
\begin{equation} \label{eq:consist1}
    \mathbf{\Tilde{S}_{b}} \cong S_b \Big\vert_{\text{on-shell}} \;\;,
\end{equation}
where $S_b$ is our proposal given in \pref{eq:boundary}. 

Additionally, even at the off-shell level, once we integrate out all the massive states following the procedure outlined in \citep{Sen:2016qap} (also see \citep{Erbin:2020eyc}), the SFT boundary term must reduce to our proposed boundary term,
\begin{equation} \label{eq:consist2}
    \mathbf{\Tilde{S}_{b}} \stackrel{\text{Wilsonian RG}}{\xrightarrow{\hspace*{2cm}}} S_b \quad .
\end{equation}

Of course, this resolution assumes the strongest form of AdS/CFT conjecture. Conversely, an explicit construction of the SFT boundary term $\mathbf{\Tilde{S}_{b}}$ that satisfies the conditions given in equations \ref{eq:consist1} and \ref{eq:consist2} would provide a piece of strong evidence in favour of the holographic principle. However, there is a glaring technical difficulty that must be overcome before such a construction is attempted.

The structure of \pref{eq:boundary} suggests that the corresponding term in SFT should also be a bi-linear in string fields, and one of the string fields must be the extra string field that is needed for the R-sector \citep{Sen:2015uaa}. However, all SFT actions are based on homotopy algebraic construction (for a review, see \citep{Erler:2019loq,Erbin:2021smf}, and bi-linear terms for any given pair of spacetime fields look naturally like $\sim \int d^{10}x \, \Phi_1(x) D \Phi_2(x)$, where $D$ is some differential operator. In fact, one can ask the same question with regards to homotopy algebraic formulation of familiar local QFTs (see \citep{Jurco:2018sby} for a review and references therein). Deciphering how to represent a total boundary term in terms of the higher products and the symplectic form (the two essential ingredients used in writing down an action in homotopy algebraic construction) will be the crucial first step in a complete construction for the boundary term in SFT.

\section{Conclusion and Outlook} \label{sec:end}

\begin{figure}[h!]
\includegraphics[scale=0.4]{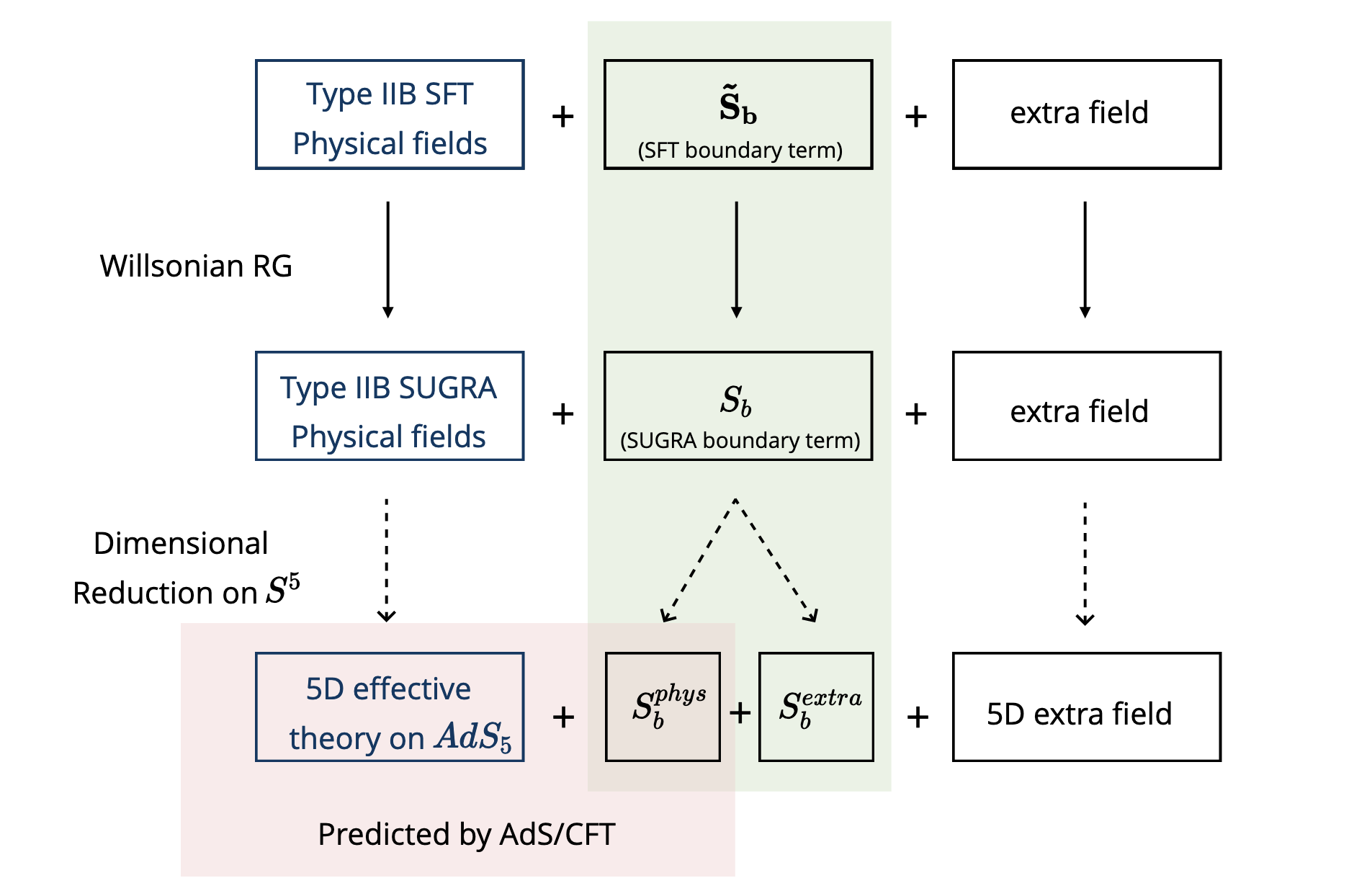}
\caption{The main findings of the paper determined the green portion such that the result is consistent with AdS/CFT. The red portion highlights the parts where AdS/CFT is assumed.}
\label{fig:1}
\end{figure}

The main result of this paper is how by adding a boundary term to Sen's action for type IIB SUGRA, we can ensure matching with the AdS/CFT prediction for the on-shell action. This, in turn, also gives powerful insight into the possible reconciliation of the full-string theory on-shell action with the AdS/CFT prediction. Figure \ref{fig:1} gives a qualitative summary of the main result. 

As discussed earlier, our proposed boundary term can also be written for any background geometry which is of the form $M_n \times M_c$, where $M_n$ is a non-compact manifold, and $M_c$ is a compact manifold. The same boundary term would lead to a non-zero on-shell action for the effective theory on $M_n$, and it will match the answer presented in \citep{Kurlyand:2022vzv}. An interesting question raised in \citep{Kurlyand:2022vzv} was if the boundary terms will have $\alpha'$-corrections once we start including the massive states of type IIB string theory. A natural answer to that, at least for our proposal, will be given by the construction of the string field theory boundary term. 

From the SUGRA side in Sen's formalism, there is no reason why the action should be supplemented by this specific boundary term. In fact, without AdS/CFT to guide us, there is no a priori reason to add \emph{any} boundary terms. This is in contrast with the approaches in \citep{Kurlyand:2022vzv,Mkrtchyan:2022xrm} where such a boundary term is seen to be expected from gauge invariance. On the other hand the result of \citep{Erler:2022agw} clearly demonstrates that for the present formulation of the SFT action we must include a boundary term to find an agreement with AdS/CFT prediction. It is possible that the SUGRA in Sen's formalism may not know about any possible constraint that uniquely fixes the boundary term. Nonetheless, assuming the technical obstacle in constructing the SFT boundary term is overcome, the expectation would be that given the conditions in equations \ref{eq:consist1} and \ref{eq:consist2}, one would hopefully end up with a unique candidate for $\mathbf{\Tilde{S}_{b}}$ in SFT. One would expect that the full SFT should know about AdS/CFT, even though that information might be non-trivially encoded. A unique boundary term for type IIB SFT on \ads which matches with AdS/CFT predictions would therefore provide crucial evidence for the strongest form of AdS/CFT conjecture.

 Another possibility is to try and make explicit connection between the Sen's formulation \citep{Sen:2015nph} and the PST formulation \citep{Pasti:1996vs,DallAgata:1997gnw} or the formulation proposed in \citep{Mkrtchyan:2022xrm}. This explicit connection will definitely help in identifying the required consistency condition for singling out the boundary term we proposed in \pref{eq:boundary}. In turn, it would also shed light into how the alternative formulations are embedded in the full string theory. 

Another interesting direction to pursue is the connection of the on-shell spacetime action for superstring with the sphere partition function of the worldsheet theory. Typically, due to the division by the volume of the $PSL(2, \mathbbm{C})$, the zero point function in worldsheet theory is naively thought to be zero. However, to be consistent with the non-vanishing of the on-shell spacetime action, the worldsheet zero point function should be suitably regularised such that it matches the spacetime answer. Important advances in this direction have already been made in the literature for open strings \citep{Eberhardt:2021ynh} and non-critical strings \citep{Mahajan:2021nsd}. It is likely that the complete construction of the boundary term in full SFT will lead to an understanding of the zero point function in the worldsheet perspective, especially since SFT is known to act as a natural regulator for the worldsheet theory \citep{Sen:2019jpm}.

\section*{Acknowledgements}
We thank Madhusudhan Raman for collaborating in the earlier stages of this work. We also thank Renann Lipinski Jusinskas and Ashoke Sen for their valuable comments on an earlier version of the manuscript and numerous discussions. AM acknowledges financial support from DST through the SERB core grant CRG/2021/000873.

\bibliography{self}

\begin{thebibliography}{10}
\providecommand{\url}[1]{\texttt{#1}}
\providecommand{\urlprefix}{URL }
\expandafter\ifx\csname urlstyle\endcsname\relax
  \providecommand{\doi}[1]{doi:\discretionary{}{}{}#1}\else
  \providecommand{\doi}{doi:\discretionary{}{}{}\begingroup
  \urlstyle{rm}\Url}\fi
\providecommand{\eprint}[2][]{\url{#2}}

\bibitem{Maldacena:1997re}
J.~M. Maldacena,
\newblock \emph{{The Large N limit of superconformal field theories and
  supergravity}},
\newblock Adv. Theor. Math. Phys. \textbf{2}, 231 (1998),
\newblock \doi{10.1023/A:1026654312961},
\newblock \eprint{hep-th/9711200}.

\bibitem{Berenstein:2002jq}
D.~E. Berenstein, J.~M. Maldacena and H.~S. Nastase,
\newblock \emph{{Strings in flat space and pp waves from N=4 superYang-Mills}},
\newblock JHEP \textbf{04}, 013 (2002),
\newblock \doi{10.1088/1126-6708/2002/04/013},
\newblock \eprint{hep-th/0202021}.

\bibitem{SCHWARZ1983301}
J.~H. Schwarz and P.~West,
\newblock \emph{Symmetries and transformations of chiral n = 2, d = 10
  supergravity},
\newblock Physics Letters B \textbf{126}(5), 301 (1983),
\newblock \doi{https://doi.org/10.1016/0370-2693(83)90168-5}.

\bibitem{HOWE1984181}
P.~Howe and P.~West,
\newblock \emph{The complete n =2, d = 10 supergravity},
\newblock Nuclear Physics B \textbf{238}(1), 181 (1984),
\newblock \doi{https://doi.org/10.1016/0550-3213(84)90472-3}.

\bibitem{SCHWARZ1983269}
J.~H. Schwarz,
\newblock \emph{Covariant field equations of chiral n = 2 d = 10 supergravity},
\newblock Nuclear Physics B \textbf{226}(2), 269 (1983),
\newblock \doi{https://doi.org/10.1016/0550-3213(83)90192-X}.

\bibitem{Kurlyand:2022vzv}
S.~A. Kurlyand and A.~A. Tseytlin,
\newblock \emph{{Type IIB supergravity action on M5\texttimes{}X5 solutions}},
\newblock Phys. Rev. D \textbf{106}(8), 086017 (2022),
\newblock \doi{10.1103/PhysRevD.106.086017},
\newblock \eprint{2206.14522}.

\bibitem{Liu:1998bu}
H.~Liu and A.~A. Tseytlin,
\newblock \emph{{D = 4 superYang-Mills, D = 5 gauged supergravity, and D = 4
  conformal supergravity}},
\newblock Nucl. Phys. B \textbf{533}, 88 (1998),
\newblock \doi{10.1016/S0550-3213(98)00443-X},
\newblock \eprint{hep-th/9804083}.

\bibitem{Henningson:1998gx}
M.~Henningson and K.~Skenderis,
\newblock \emph{{The Holographic Weyl anomaly}},
\newblock JHEP \textbf{07}, 023 (1998),
\newblock \doi{10.1088/1126-6708/1998/07/023},
\newblock \eprint{hep-th/9806087}.

\bibitem{Gubser:1998vd}
S.~S. Gubser,
\newblock \emph{{Einstein manifolds and conformal field theories}},
\newblock Phys. Rev. D \textbf{59}, 025006 (1999),
\newblock \doi{10.1103/PhysRevD.59.025006},
\newblock \eprint{hep-th/9807164}.

\bibitem{Blau:1999vz}
M.~Blau, K.~S. Narain and E.~Gava,
\newblock \emph{{On subleading contributions to the AdS / CFT trace anomaly}},
\newblock JHEP \textbf{09}, 018 (1999),
\newblock \doi{10.1088/1126-6708/1999/09/018},
\newblock \eprint{hep-th/9904179}.

\bibitem{Burgess:1999vb}
C.~P. Burgess, N.~R. Constable and R.~C. Myers,
\newblock \emph{{The Free energy of N=4 superYang-Mills and the AdS / CFT
  correspondence}},
\newblock JHEP \textbf{08}, 017 (1999),
\newblock \doi{10.1088/1126-6708/1999/08/017},
\newblock \eprint{hep-th/9907188}.

\bibitem{Russo:2012ay}
J.~G. Russo and K.~Zarembo,
\newblock \emph{{Large N Limit of N=2 SU(N) Gauge Theories from Localization}},
\newblock JHEP \textbf{10}, 082 (2012),
\newblock \doi{10.1007/JHEP10(2012)082},
\newblock \eprint{1207.3806}.

\bibitem{Sen:2015uaa}
A.~Sen,
\newblock \emph{{BV Master Action for Heterotic and Type II String Field
  Theories}},
\newblock JHEP \textbf{02}, 087 (2016),
\newblock \doi{10.1007/JHEP02(2016)087},
\newblock \eprint{1508.05387}.

\bibitem{Sen:2015nph}
A.~Sen,
\newblock \emph{{Covariant Action for Type IIB Supergravity}},
\newblock JHEP \textbf{07}, 017 (2016),
\newblock \doi{10.1007/JHEP07(2016)017},
\newblock \eprint{1511.08220}.

\bibitem{Mkrtchyan:2022xrm}
K.~Mkrtchyan and F.~Valach,
\newblock \emph{{Actions for type II supergravities and democracy}}  (2022),
\newblock \eprint{2207.00626}.

\bibitem{Erler:2022agw}
T.~Erler,
\newblock \emph{{The closed string field theory action vanishes}},
\newblock JHEP \textbf{10}, 055 (2022),
\newblock \doi{10.1007/JHEP10(2022)055},
\newblock \eprint{2204.12863}.

\bibitem{Sen:2019qit}
A.~Sen,
\newblock \emph{{Self-dual forms: Action, Hamiltonian and Compactification}},
\newblock J. Phys. A \textbf{53}(8), 084002 (2020),
\newblock \doi{10.1088/1751-8121/ab5423},
\newblock \eprint{1903.12196}.

\bibitem{Lambert:2019diy}
N.~Lambert,
\newblock \emph{{(2,0) Lagrangian Structures}},
\newblock Phys. Lett. B \textbf{798}, 134948 (2019),
\newblock \doi{10.1016/j.physletb.2019.134948},
\newblock \eprint{1908.10752}.

\bibitem{Andriolo:2020ykk}
E.~Andriolo, N.~Lambert and C.~Papageorgakis,
\newblock \emph{{Geometrical Aspects of An Abelian (2,0) Action}},
\newblock JHEP \textbf{04}, 200 (2020),
\newblock \doi{10.1007/JHEP04(2020)200},
\newblock \eprint{2003.10567}.

\bibitem{Gustavsson:2020ugb}
A.~Gustavsson,
\newblock \emph{{A nonabelian M5 brane Lagrangian in a supergravity
  background}},
\newblock JHEP \textbf{10}, 001 (2020),
\newblock \doi{10.1007/JHEP10(2020)001},
\newblock \eprint{2006.07557}.

\bibitem{Chakrabarti:2020dhv}
S.~Chakrabarti, D.~Gupta, A.~Manna and M.~Raman,
\newblock \emph{{Irrelevant deformations of chiral bosons}},
\newblock JHEP \textbf{02}, 028 (2021),
\newblock \doi{10.1007/JHEP02(2021)028},
\newblock \eprint{2011.06352}.

\bibitem{Andriolo:2021gen}
E.~Andriolo, N.~Lambert, T.~Orchard and C.~Papageorgakis,
\newblock \emph{{A path integral for the chiral-form partition function}},
\newblock JHEP \textbf{04}, 115 (2022),
\newblock \doi{10.1007/JHEP04(2022)115},
\newblock \eprint{2112.00040}.

\bibitem{Rist:2020uaa}
D.~Rist, C.~Saemann and M.~van~der Worp,
\newblock \emph{{Towards an M5-brane model. Part III. Self-duality from
  additional trivial fields}},
\newblock JHEP \textbf{06}, 036 (2021),
\newblock \doi{10.1007/JHEP06(2021)036},
\newblock \eprint{2012.09253}.

\bibitem{Cremonini:2020skt}
C.~A. Cremonini and P.~A. Grassi,
\newblock \emph{{Self-dual forms in supergeometry I: The chiral boson}},
\newblock Nucl. Phys. B \textbf{973}, 115596 (2021),
\newblock \doi{10.1016/j.nuclphysb.2021.115596},
\newblock \eprint{2012.10243}.

\bibitem{Chakrabarti:2022lnn}
S.~Chakrabarti, A.~Manna and M.~Raman,
\newblock \emph{{Renormalization in $\text{T}\overline{\text{T}}$-deformed
  nonintegrable theories}},
\newblock Phys. Rev. D \textbf{105}(10), 106025 (2022),
\newblock \doi{10.1103/PhysRevD.105.106025},
\newblock \eprint{2204.03385}.

\bibitem{Andrianopoli:2022bzr}
L.~Andrianopoli, C.~A. Cremonini, R.~D'Auria, P.~A. Grassi, R.~Matrecano,
  R.~Noris, L.~Ravera and M.~Trigiante,
\newblock \emph{{M5-brane in the superspace approach}},
\newblock Phys. Rev. D \textbf{106}(2), 026010 (2022),
\newblock \doi{10.1103/PhysRevD.106.026010},
\newblock \eprint{2206.06388}.

\bibitem{Pasti:1996vs}
P.~Pasti, D.~P. Sorokin and M.~Tonin,
\newblock \emph{{On Lorentz invariant actions for chiral p forms}},
\newblock Phys. Rev. D \textbf{55}, 6292 (1997),
\newblock \doi{10.1103/PhysRevD.55.6292},
\newblock \eprint{hep-th/9611100}.

\bibitem{DallAgata:1997gnw}
G.~Dall'Agata, K.~Lechner and D.~P. Sorokin,
\newblock \emph{{Covariant actions for the bosonic sector of d = 10 IIB
  supergravity}},
\newblock Class. Quant. Grav. \textbf{14}, L195 (1997),
\newblock \doi{10.1088/0264-9381/14/12/003},
\newblock \eprint{hep-th/9707044}.

\bibitem{deLacroix:2017lif}
C.~de~Lacroix, H.~Erbin, S.~P. Kashyap, A.~Sen and M.~Verma,
\newblock \emph{{Closed Superstring Field Theory and its Applications}},
\newblock Int. J. Mod. Phys. A \textbf{32}(28n29), 1730021 (2017),
\newblock \doi{10.1142/S0217751X17300216},
\newblock \eprint{1703.06410}.

\bibitem{Erler:2019loq}
T.~Erler,
\newblock \emph{{Four Lectures on Closed String Field Theory}},
\newblock Phys. Rept. \textbf{851}, 1 (2020),
\newblock \doi{10.1016/j.physrep.2020.01.003},
\newblock \eprint{1905.06785}.

\bibitem{Erbin:2021smf}
H.~Erbin,
\newblock \emph{{String Field Theory: A Modern Introduction}}, vol. 980 of
  \emph{Lecture Notes in Physics},
\newblock ISBN 978-3-030-65320-0, 978-3-030-65321-7,
\newblock \doi{10.1007/978-3-030-65321-7} (2021).

\bibitem{Fradkin:1984pq}
E.~S. Fradkin and A.~A. Tseytlin,
\newblock \emph{{Effective Field Theory from Quantized Strings}},
\newblock Phys. Lett. B \textbf{158}, 316 (1985),
\newblock \doi{10.1016/0370-2693(85)91190-6}.

\bibitem{Fradkin:1985ys}
E.~S. Fradkin and A.~A. Tseytlin,
\newblock \emph{{Quantum String Theory Effective Action}},
\newblock Nucl. Phys. B \textbf{261}, 1 (1985),
\newblock \doi{10.1016/0550-3213(85)90559-0},
\newblock [Erratum: Nucl.Phys.B 269, 745--745 (1986)].

\bibitem{Callan:1986jb}
C.~G. Callan, Jr., I.~R. Klebanov and M.~J. Perry,
\newblock \emph{{String Theory Effective Actions}},
\newblock Nucl. Phys. B \textbf{278}, 78 (1986),
\newblock \doi{10.1016/0550-3213(86)90107-0}.

\bibitem{Tseytlin:1988tv}
A.~A. Tseytlin,
\newblock \emph{{Mobius Infinity Subtraction and Effective Action in $\sigma$
  Model Approach to Closed String Theory}},
\newblock Phys. Lett. B \textbf{208}, 221 (1988),
\newblock \doi{10.1016/0370-2693(88)90421-2}.

\bibitem{Tseytlin:1988rr}
A.~A. Tseytlin,
\newblock \emph{{SIGMA MODEL APPROACH TO STRING THEORY}},
\newblock Int. J. Mod. Phys. A \textbf{4}, 1257 (1989),
\newblock \doi{10.1142/S0217751X8900056X}.

\bibitem{Susskind:1994sm}
L.~Susskind and J.~Uglum,
\newblock \emph{{Black hole entropy in canonical quantum gravity and
  superstring theory}},
\newblock Phys. Rev. D \textbf{50}, 2700 (1994),
\newblock \doi{10.1103/PhysRevD.50.2700},
\newblock \eprint{hep-th/9401070}.

\bibitem{Chen:2021dsw}
Y.~Chen, J.~Maldacena and E.~Witten,
\newblock \emph{{On the black hole/string transition}}  (2021),
\newblock \eprint{2109.08563}.

\bibitem{Eberhardt:2021ynh}
L.~Eberhardt and S.~Pal,
\newblock \emph{{The disk partition function in string theory}},
\newblock JHEP \textbf{08}, 026 (2021),
\newblock \doi{10.1007/JHEP08(2021)026},
\newblock \eprint{2105.08726}.

\bibitem{Mahajan:2021nsd}
R.~Mahajan, D.~Stanford and C.~Yan,
\newblock \emph{{Sphere and disk partition functions in Liouville and in matrix
  integrals}},
\newblock JHEP \textbf{07}, 132 (2022),
\newblock \doi{10.1007/JHEP07(2022)132},
\newblock \eprint{2107.01172}.

\bibitem{Sen:2016qap}
A.~Sen,
\newblock \emph{{Wilsonian Effective Action of Superstring Theory}},
\newblock JHEP \textbf{01}, 108 (2017),
\newblock \doi{10.1007/JHEP01(2017)108},
\newblock \eprint{1609.00459}.

\bibitem{Erbin:2020eyc}
H.~Erbin, C.~Maccaferri, M.~Schnabl and J.~Vo\v{s}mera,
\newblock \emph{{Classical algebraic structures in string theory effective
  actions}},
\newblock JHEP \textbf{11}, 123 (2020),
\newblock \doi{10.1007/JHEP11(2020)123},
\newblock \eprint{2006.16270}.

\bibitem{Jurco:2018sby}
B.~Jur\v{c}o, L.~Raspollini, C.~S\"amann and M.~Wolf,
\newblock \emph{{$L_\infty$-Algebras of Classical Field Theories and the
  Batalin-Vilkovisky Formalism}},
\newblock Fortsch. Phys. \textbf{67}(7), 1900025 (2019),
\newblock \doi{10.1002/prop.201900025},
\newblock \eprint{1809.09899}.

\bibitem{Sen:2019jpm}
A.~Sen,
\newblock \emph{{String Field Theory as World-sheet UV Regulator}},
\newblock JHEP \textbf{10}, 119 (2019),
\newblock \doi{10.1007/JHEP10(2019)119},
\newblock \eprint{1902.00263}.

\end{thebibliography}

\end{document}